\documentclass[twocolumn,showpacs,preprintnumbers,amsmath,amssymb]{revtex4}
\usepackage{graphicx}% Include figure files
\usepackage{dcolumn}% Align table columns on decimal point
\usepackage{bm}% bold math
\begin{document}
\preprint{APS/123-QED}
\title{Comparative study of high-spin isomers in semi-magic $Z$=50 isotopic and $N$=82 isotonic chains}% Force line breaks with \\
\author{Bhoomika Maheshwari}
\email{bhoomika.physics@gmail.com}
 %\altaffiliation[Also at ]{Department of Physics, Indian Institute of Technology, Roorkee, India.}%Lines break automatically or can be forced with \\
\author{Ashok Kumar Jain}%
%\affiliation{%
%\author{P.C. Srivastava}
%\affiliation{
\affiliation{Department of Physics, Indian Institute of Technology, Roorkee 247667, India.}
\date{\today}% It is always \today, today,
\begin{abstract}
%An article usually includes an abstract, a concise summary of the work
%covered at length in the main body of the article. It is used for
%secondary publications and for information retrieval purposes. Valid
%PACS numbers may be entered using the \verb+\pacs{#1}+ command.

We present a comparative study of the high-spin ${11/2}^-$, ${10}^+$ and ${27/2}^-$ nuclear isomers, observed  commonly in the full semi-magic $Z=50$ isotopic and $N=82$ isotonic chains, which are found to display nearly identical systematics in excitation energy and half-lives. A constant and particle number independent energy gap of $\sim$ 4 MeV between the $0^+$ ground states and ${10}^+$ isomers and, the ${11/2}^-$ and ${27/2}^-$ isomers exists before the mid-shell, which declines near the mid-shell, and then becomes constant at $\sim$ 3 MeV after the mid-shell. Large scale shell model calculations are carried out for the full chains of Sn-isotopes and $N=82$ isotones, which reproduce the observed energy systematics reasonably well. The role of seniority, generalized seniority and alignment in the high-spin isomeric states as well as in the states populating the high spin isomers has been discussed. The empirical features suggest that the generalized seniority, known to be valid in the low lying states of even-even nuclei before the mid-shell, remains good up to the ${10}^+$ isomeric states; we further extend this interpretation to the odd-A nuclei before the mid-shell, particularly the ${27/2}^-$ isomers. The seniority becomes good after the mid-shell. The change in energy gap around the mid-shell, therefore, marks a transition from the generalized seniority regime to the seniority regime. From the alignment considerations, the high-spin isomers after the mid-shell emerge as maximally aligned decoupled states. We also note that the similar features of the isomeric half-lives in both the chains may be understood in terms of their similar seniorities and configurations. An overview of the $0^+$ to ${10}^+$ yrast states in the full chain of even-even Sn-isotopes has also been presented. We also make some predictions of possible new isomers based on the empirical systematics.

\end{abstract}
\maketitle
\section{\label{sec:level1}Introduction}
Nuclear isomers, which are metastable states of nuclei, may be regarded as a separate class of nuclei and their study has recently gained prominence because of many reasons, both fundamental as well as applied~\cite{walker99}. The experimental data on isomers have been growing significantly with the modern instrumentation and new radioactive beams. Our “Atlas of Nuclear Isomers” lists around 2450 nuclear isomers with a half-life cut-off at 10 ns~\cite{jain15}. This data set reveals a variety of systematics and novel features across the whole nuclear landscape~\cite{jain14}. Nuclear isomers are generally classified into four types: the spin isomers, the K isomers, the shape isomers and the fission isomers~\cite{walker99}. Besides these, a fifth type known as the seniority isomers also exists, which mostly occurs in the semi-magic nuclei. 

Racah had originally introduced the concept of seniority in the atomic context~\cite{racah42}. Later on, this concept was extended to nuclei and seniority was defined as the number of unpaired nucleons in a nucleus. It has been shown that the seniority quantum number remains conserved in the states arising from identical nucleons in a single-j orbital for j $\leq$ 7/2~\cite{talmi93, lawson80}. But as we go to the higher-j orbitals, seniority may remain only partially conserved~\cite{talmi93, lawson80, talmi03}. A detailed discussion of the partial seniority conservation in the high-j orbitals and the effective interactions may be found in~\cite{talmi03}. Even then, we find that the high-j intruder orbitals having j $\geq$ 7/2 appear to play an important role in seniority becoming a good quantum number in the semi-magic nuclei, which in turn leads to the occurrence of the seniority isomers due to the seniority selection rules~\cite{talmi93, lawson80, talmi03, isacker10, isacker11}. There are some cases when the identical nucleons occupy several-j orbitals and lead to configuration mixing, and still show some empirical features of the seniority scheme. These were understood in terms of the generalized seniority~\cite{talmi71}. The validity of the generalized seniority scheme has been tested for the low lying states in the even-even lighter Sn-isotopes~\cite{sandulescu97}, and in the $N=82$ isotones before the mid-shell~\cite{scholten83}.

The present work focuses on the high spin isomers in the $Z=50$ isotopes and the $N=82$ isotones, which are unique as they span the same valence nucleon space $50-82$. To obtain the complete empirical systematics in this region, we have relaxed the cut-off limit of isomeric half-life $\geq$ 10 ns in the present study, and considered all the known yrast ${11/2}^-$, ${10}^+$ and ${27/2}^-$ isomers, for both the chains. This allows us to compare the behavior of the high-spin ${11/2}^-$, ${10}^+$ and ${27/2}^-$ isomers in the Sn-isotopes $(N=54$ to $81)$ approaching the neutron-drip line with the isomers in the $N=82$ isotones $(Z=51$ to $73)$ approaching the proton-drip line.

In section ~\ref{sec:level2}, we compare the experimental energy systematics of the ${11/2}^-$, ${10}^+$ and ${27/2}^-$ isomers throughout the $Z=50$ and $N=82$ chains. To study their structure, we have carried out large scale shell model calculations for both the chains. The details of these calculations and the calculated systematics along with their effective single particle energies (ESPE) are presented in section ~\ref{sec:level3}. The empirical behavior along with the shell model calculations allows us to infer the seniority and generalized seniority of the isomers, many of them in agreement with earlier studies. Section ~\ref{sec:level4} focuses on their configurations and the wave functions to fix the seniority of these isomers in both the chains. We discuss the seniority, generalized seniority and alignment properties of the isomers as well as the states involved in the population of the isomers. We, further, present a complete empirical picture of the even-even Sn-isotopes in view of our observations so far. We also provide a qualitative understanding of the half-life systematics in terms of the seniority scheme in section ~\ref{sec:level5}. Such comparative studies allow us to bring out the similarities and differences that may arise when protons are replaced by neutrons in the same set of orbitals for the $Z=50$ isotopic and $N=82$ isotonic chains. We also predict the existence of some isomers and their properties based on our systematic studies in section ~\ref{sec:level6}. Section~\ref{sec:level7} summarizes the present work.
%%%%%%%%%%%%%%%%%%%%%%%%%%%%%%%%%%%%%%%%%%%%%%%%%%%%%%%%%%%%%%%%%%%%%%%%%%%%%%%%%%%
\begin{figure}[!ht]
\includegraphics[width=8cm,height=8cm]{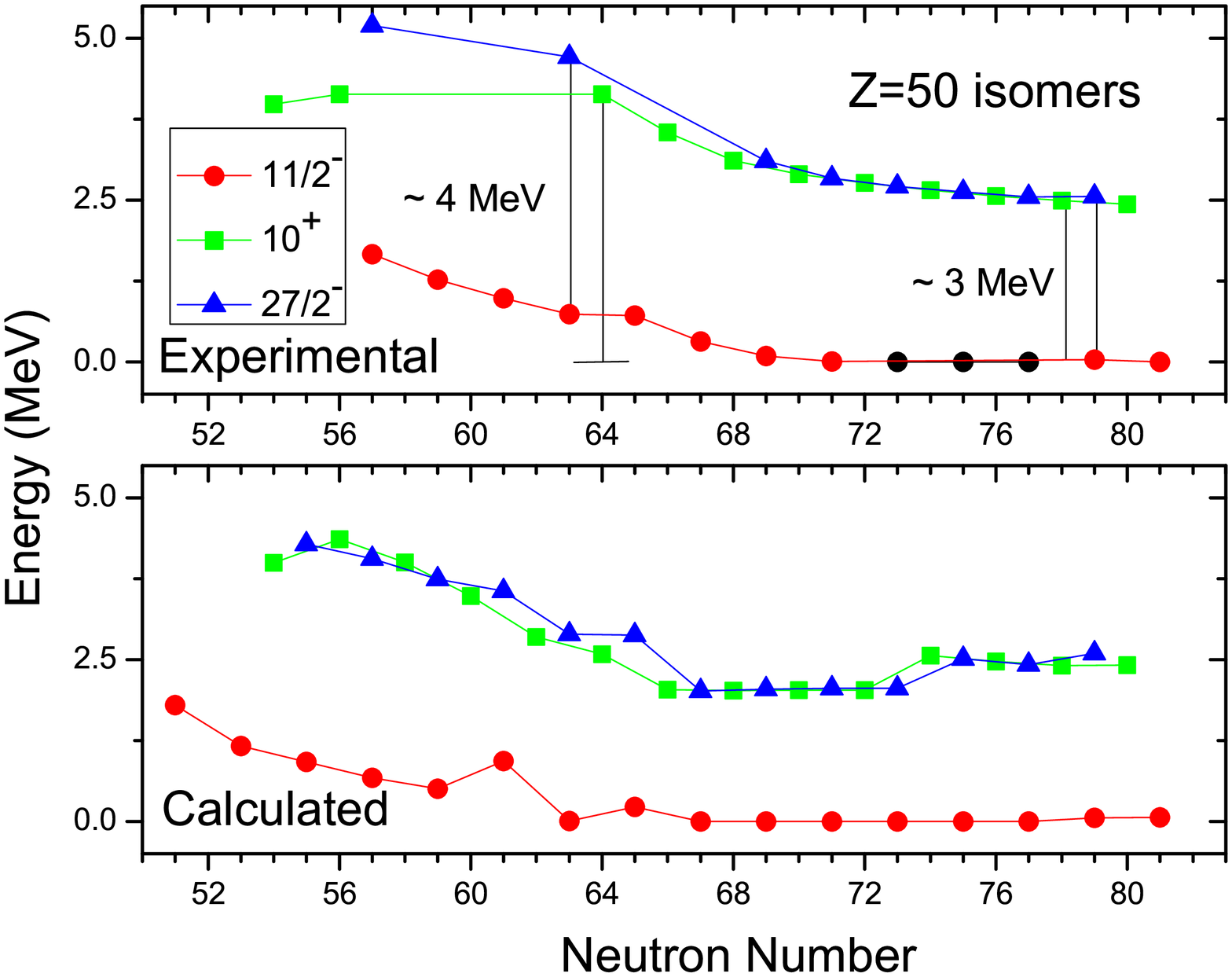}% Here is how to import EPS art
\caption{\label{fig:fig1}(Color online) Comparison of the calculated and the experimental excitation energy systematics of isomeric states in Sn-isotopes. Black circles represent the ground states for ${11/2}^-$.}
\end{figure}
%%%%%%%%%%%%%%%%%%%%%%%%%%%%%%%%%%%%%%%%%%%%%%%%%%%%%%%%%%%%%%%%%%%%%%%%%%%%%%%%%%%%%
%%%%%%%%%%%%%%%%%%%%%%%%%%%%%%%%%%%%%%%%%%%%%%%%%%%%%%%%%%%%%%%%%%%%%%%%%%%%%%%%%%%%%%%%%%%%%
\begin{figure}[!ht]
\includegraphics[width=8cm,height=8cm]{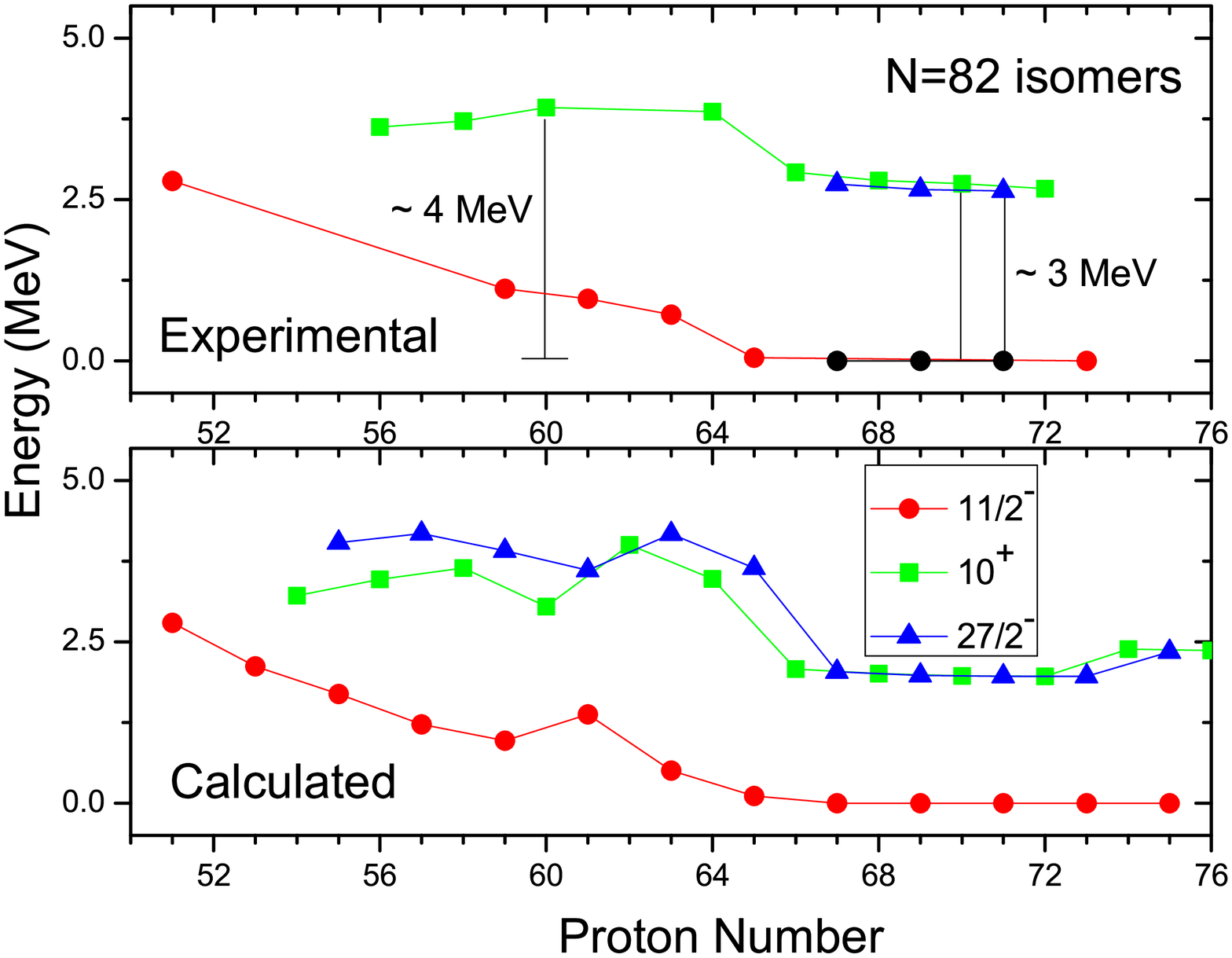}% Here is how to import EPS art
\caption{\label{fig:fig2}(Color online) Comparison of the calculated and the experimental excitation energy systematics of isomeric states for $N=82$ isotones. Black circles represent the ground states for ${11/2}^-$.}
\end{figure}
%%%%%%%%%%%%%%%%%%%%%%%%%%%%%%%%%%%%%%%%%%%%%%%%%%%%%%%%%%%%%%%%%%%%%%%%%%%%%%%%%%%
\begin{figure}[!ht]
\includegraphics[width=8cm,height=7cm]{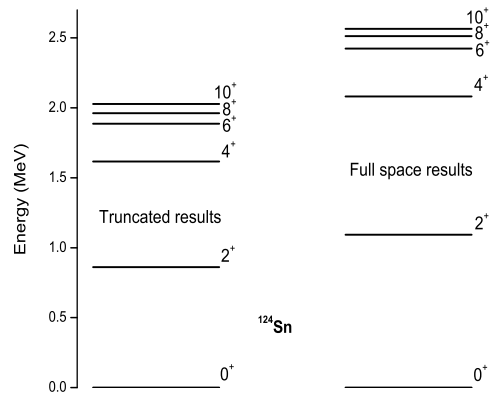}% Here is how to import EPS art
\caption{\label{fig:fig3} A comparison of full valence space results with the truncated ones in the case of $^{124}$Sn. The truncated calculations have been done by freezing the g$_{7/2}$ orbital.}
\end{figure}
%%%%%%%%%%%%%%%%%%%%%%%%%%%%%%%%%%%%%%%%%%%%%%%%%%%%%%%%%%%%%%%%%%%%%%%%%%%%%%%%%%%%%
\begin{figure}[!ht]
\includegraphics[width=8cm,height=7cm]{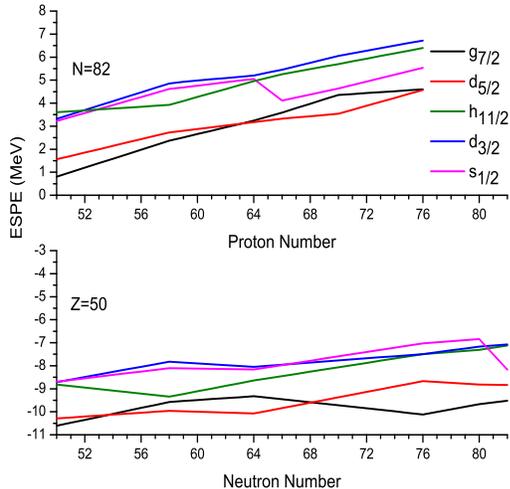}% Here is how to import EPS art
\caption{\label{fig:fig4}(Color online) Effective single particle energies for the $Z=50$ isotopes and the $N=82$ isotones.}
\end{figure}
%%%%%%%%%%%%%%%%%%%%%%%%%%%%%%%%%%%%%%%%%%%%%%%%%%%%%%%%%%%%%%%%%%%%%%%%%%%%%%%%%%%%%
%%%%%%%%%%%%%%%%%%%%%%%%%%%%%%%%%%%%%%%%%%%%%%%%%%%%%%%%%%%%%%%%%%%%%%%%%%%%%%%%%%%%%
\begin{figure}[!ht]
\includegraphics[width=8cm,height=7cm]{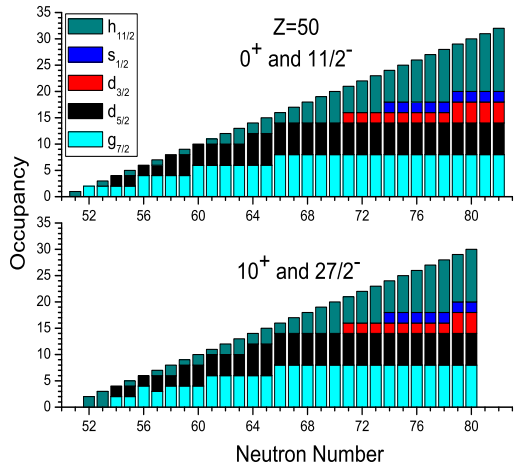}% Here is how to import EPS art
\caption{\label{fig:fig5}(Color online) Variation of occupancy corresponding to the maximum partition for the $Z$=50 isotopes.}
\end{figure}
%%%%%%%%%%%%%%%%%%%%%%%%%%%%%%%%%%%%%%%%%%%%%%%%%%%%%%%%%%%%%%%%%%%%%%%%%%%%%%%%%%%%%%%
%%%%%%%%%%%%%%%%%%%%%%%%%%%%%%%%%%%%%%%%%%%%%%%%%%%%%%%%%%%%%%%%%%%%%%%%%%%%%%%%%%%%%
\begin{figure}[!ht]
\includegraphics[width=8cm,height=7cm]{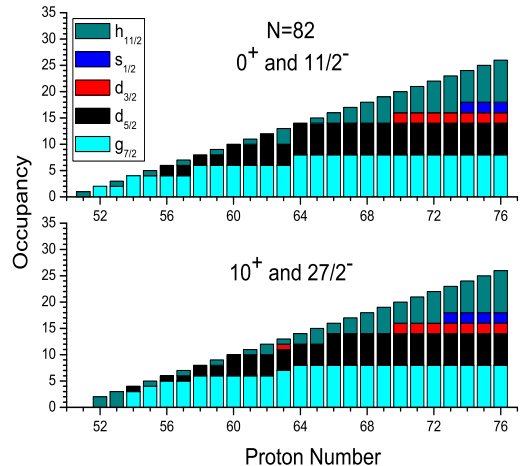}% Here is how to import EPS art
\caption{\label{fig:fig6}(Color online) Same as Fig.~\ref{fig:fig5} but for the $N$=82 isotones.}
\end{figure}
%%%%%%%%%%%%%%%%%%%%%%%%%%%%%%%%%%%%%%%%%%%%%%%%%%%%%%%%%%%%%%%%%%%%%%%%%%%%%%%%%%%%%

\section{\label{sec:level2}Experimental systematics of the high-spin semi-magic isomers}

The level schemes of the $^{119-130}$Sn-isotopes have been studied by using the reactions induced by light ions, deep inelastic reactions, or fission fragment studies by several researchers~\cite{daly80, fogelberg81, daly86, lunardi87, broda92, mayer94, daly95, pinston00, zhang00, lozeva08}. Many isomer systematics have been identified for $N>64$ Sn-isotopes, and the isomeric states ${10}^+$ and ${27/2}^-$ have been characterized as seniority $\it{v}$=2 and $\it{v}$=3 states in these studies. Pietri et al.~\cite{pietri11} recently identified and confirmed the high-spin and high-seniority $\it{v}$=4, ${15}^-$ isomeric state in $^{128}$Sn. More recently Astier et al.~\cite{astier132, astier125} reported detailed high-spin level schemes in the $^{119-126}$Sn-isotopes by using the binary fission fragmentation induced by heavy ions. Iskra et al.~\cite{iskra14} have also focused on high-seniority states in neutron-rich, even-even Sn-isotopes. It may be noted that there exists some deformed collective states giving rise to a full or a part of rotational band in the even-even light mass Sn-isotopes with $A=110-118$, interpreted as 2p-2h proton configuration~\cite{bron79, poelgeest80, harada88, savelius98, gableske01, wolinska05, wang10}. But the ${10}^+$ yrast isomeric states discussed in the present paper are not part of any rotational structure~\cite{fotiades11}. More recently, the studies on Sn-isotopes have been pushed much beyond the $N=82$ shell closure and isomers in the $N=86-88$ Sn-isotopes have been populated by Simpson et al.~\cite{simpson14} which shed a new light on the effective interaction in n-rich nuclei [36]. 

The ${10}^+$ and ${27/2}^-$ isomers have also been identified, in the $N=82$ isotonic chain from $Z=66$, Dy to $Z=72$, Hf, as seniority $\it{v}$=2 and $\it{v}$=3 isomers coming from the $h_{11/2}$ proton orbital~\cite{mcneill89}. Recently, the high-spin structure of five $N=82$ isotones with $Z=54-58$ has also been reported by Astier et al.~\cite{astier126}. The ${10}^+$ isomers have been described as broken pairs of protons from the $g_{7/2}$ and $d_{5/2}$ orbitals in the even-mass isotones, before the mid-shell~\cite{astier126}.

We plot the complete systematics of the measured excitation energies of the ${11/2}^-$, ${10}^+$ and ${27/2}^-$ isomers for the $Z=50$ isotopic and the $N=82$ isotonic chains in the top panels of Fig.~\ref{fig:fig1} and Fig.~\ref{fig:fig2} respectively. It may be noted that the same valence orbitals are involved in both the chains. While the neutrons occupy these orbitals in the $Z=50$ isomers, the protons take over the role in the $N=82$ isomers. We find that all the main features observed in the $Z=50$ isotopic chain are also present in the $N=82$ isotonic chain and both appear to be nearly identical to each other. 

The available experimental data from the nucleon number 51 to 58 for both the chains of isomers are limited. Yet we can see that the ${11/2}^-$ isomeric state exhibits a gradual decline in energy from $\sim$ 1.5 MeV to $\sim$ 0 MeV as the valence nucleon number crosses the mid-shell at 59. With further increase in the valence nucleon number, the ${11/2}^-$ isomeric state briefly becomes the ground state for $N=73-77$ and $Z=67-71$, and again becomes an isomeric state at $N>77$ and $Z>71$, for the $Z=50$ isotopic and $N=82$ isotonic chains, respectively. 

The relative energy gap of $\sim$ 4 MeV before the mid-shell, between the ${11/2}^-$ and ${27/2}^-$ isomers, also reduces to $\sim$ 3 MeV after the mid-shell at N or, $Z=66$. In a similar way, the ${10}^+$ isomeric state lies at $\sim$ 4 MeV from the $0^+$ ground state before the mid-shell, and comes down to $\sim$ 3 MeV after the mid-shell. Thus, the energy gap is constant and particle number independent before and after the mid-shell, which is a well known signature of nearly good seniority ~\cite{talmi93, lawson80, talmi03}. The ${10}^+$ and the ${27/2}^-$ isomers, belonging to the even-even and even-odd nuclei respectively, are seen to follow each other very closely throughout the chains, if one puts the $0^+$ and ${11/2}^-$ states on equal footing. This suggests that the configurations for the ${10}^+$ and the ${27/2}^-$ isomers should be very similar in nature, before and after the mid-shell.

%%%%%%%%%%%%%%%%%%%%%%%%%%%%%%%%%%%%%%%%%%%%%%%%%%%%%%%%%%%%%%%%%%%%%%%%%%%%%%%%%%%%%%%%%%%%
\begin{table*}[htb]
\caption{\label{tab:table1}Seniority assignments of the isomeric states ${11/2}^-$, ${10}^+$ and ${27/2}^-$ in the Sn-isotopic chain, where the unpaired nucleons in the respective orbitals are listed as configuration.}
\begin{ruledtabular}
\begin{tabular}{c c c c c c c c}
${ }$& \multicolumn{2}{c}{${10}^+$} & {} & \multicolumn{2}{c}{${11/2}^-$} & \multicolumn{2}{c}{${27/2}^-$} \\
\hline\hline
Isotope   &  Configuration &  Seniority & Isotope & Configuration &  Seniority & Configuration &  Seniority \\
\hline
&&&&&&&\\
$^{102}$Sn  & $h_{11/2}^2$ & ${2}$ & $^{103}$Sn & $h_{11/2}^1$& ${1}$  & $h_{11/2}^3$& ${3}$\\
$^{104}$Sn  & $(g_{7/2} d_{5/2})^4$ & ${4}$ & $^{105}$Sn  & $h_{11/2}^1$& ${1}$ & $(g_{7/2} d_{5/2})^4$ $h_{11/2}^1$ &${5}$\\
$^{106}$Sn  & $(g_{7/2} d_{5/2})^4$ & ${4}$ & $^{107}$Sn  & $h_{11/2}^1$& ${1}$ & $(g_{7/2} d_{5/2})^4$ $h_{11/2}^1$ &${5}$\\
$^{108}$Sn  & $h_{11/2}^2$ & ${2}$ & $^{109}$Sn   & $h_{11/2}^1$ & ${1}$ & $(g_{7/2} d_{5/2})^4$ $h_{11/2}^1$ &${5}$\\
$^{110}$Sn  & $h_{11/2}^2$ & ${2}$ & $^{111}$Sn  & $h_{11/2}^1$& ${1}$  & $(g_{7/2} d_{5/2})^4$ $h_{11/2}^1$ &${5}$\\
$^{112}$Sn & $h_{11/2}^2$ & ${2}$ & $^{113}$Sn  & $h_{11/2}^1$ & ${1}$  & $h_{11/2}^3$ &${3}$\\
$^{114}$Sn & $h_{11/2}^2$ & ${2}$ & $^{115}$Sn  & $h_{11/2}^1$ & ${1}$  & $h_{11/2}^3$&${3}$ \\
\end{tabular}
\end{ruledtabular}
\end{table*}
%%%%%%%%%%%%%%%%%%%%%%%%%%%%%%%%%%%%%%%%%%%%%%%%%%%%%%%%%%%%%%%%%%%%%%%%%%%%%%%%%%%%%%%%%%%%%%
%%%%%%%%%%%%%%%%%%%%%%%%%%%%%%%%%%%%%%%%%%%%%%%%%%%%%%%%%%%%%%%%%%%%%%%%%%%%%%%%%%%%%%%%%%%%%%
\begin{table*}[htb]
\caption{\label{tab:table2}Seniority assignments of the isomeric states ${11/2}^-$, ${10}^+$ and ${27/2}^-$ in the $N=82$ isotonic chain, where the unpaired nucleons in the respective orbitals are listed as configuration.}
\begin{ruledtabular}
\begin{tabular}{c c c c c c c c}
${ }$& \multicolumn{2}{c}{${10}^+$} & {} & \multicolumn{2}{c}{${11/2}^-$} & \multicolumn{2}{c}{${27/2}^-$} \\
\hline\hline
Isotone  &  Configuration &  Seniority & Isotone & Configuration & Seniority & Configuration &  Seniority\\
\hline
&&&&&&&\\
$^{134}$Te  & $h_{11/2}^2$ & ${2}$& $^{135}$I & $h_{11/2}^1$ & ${1}$ & $h_{11/2}^3$ & ${3}$ \\
$^{136}$Xe  & $(g_{7/2} d_{5/2})^4$ & ${4}$& $^{137}$Cs & $h_{11/2}^1$ & ${1}$ & $(g_{7/2} d_{5/2})^4$ $h_{11/2}^1$ & ${5}$ \\
$^{138}$Ba  & $(g_{7/2} d_{5/2})^4$ & ${4}$ & $^{139}$La & $h_{11/2}^1$ & ${1}$ & $(g_{7/2} d_{5/2})^4$ $h_{11/2}^1$ & ${5}$ \\
$^{140}$Ce  & $(g_{7/2} d_{5/2})^4$ & ${4}$ & $^{141}$Pr & $h_{11/2}^1$ & ${1}$ & $(g_{7/2} d_{5/2})^4$ $h_{11/2}^1$ & ${5}$ \\
$^{142}$Nd   & $(g_{7/2} d_{5/2})^4$ & ${4}$ & $^{143}$Pm & $h_{11/2}^1$ & ${1}$ & $(g_{7/2} d_{5/2})^4$ $h_{11/2}^1$ & ${5}$ \\
$^{144}$Sm & $(g_{7/2} d_{5/2})^4$ & ${4}$ & $^{145}$Eu & $h_{11/2}^1$ & ${1}$ & $(g_{7/2} d_{5/2})^4$ $h_{11/2}^1$ & ${5}$ \\
$^{146}$Gd & $h_{11/2}^2$ & ${2}$ & $^{147}$Tb & $h_{11/2}^1$ & ${1}$ & $h_{11/2}^3$ & ${3}$ \\
\end{tabular}
\end{ruledtabular}
\end{table*}
%%%%%%%%%%%%%%%%%%%%%%%%%%%%%%%%%%%%%%%%%%%%%%%%%%%%%%%%%%%%%%%%%%%%%%%%%%%%%%%%%%%%%%%%%%%%
\begin{table*}[htb]
\caption{\label{tab:table3}Comparison of the experimentally measured $\Delta E^{2^+}_{0^+}$ and $ \Delta E^{{15/2}^-}_{{11/2}^-}$ $\gamma-$transitions in even-even and odd-A Sn-isotopes for $N$ $\ge$ 64. Also, compared are the $\Delta E^{12^+}_{10^+}$ and $ \Delta E^{{31/2}^-}_{{27/2}^-}$ $\gamma-$transitions involving states which decay to the ${10}^+$, ${27/2}^-$ isomers. $R(15:2)$  is defined as $\Delta E^{{15/2}^-}_{{11/2}^-}$ $/ \Delta E^{2^+}_{0^+}$ and $R(31:12)$ is defined as $\Delta E^{{31/2}^-}_{{27/2}^-}$ $/ \Delta E^{12^+}_{10^+}$. All the energies are in MeV.}
\begin{ruledtabular}
\begin{tabular}{c c c c c c c c}
Isotope & $ \Delta E^{2^+}_{0^+} $ & $\Delta E^{12^+}_{10^+}$ & Isotope & $\Delta E^{{15/2}^-}_{{11/2}^-}$ & $\Delta E^{{31/2}^-}_{{27/2}^-}$ & $R(15:2)$ & $R(31:12)$ \\
\hline
&&&&\\
$^{112}$Sn & ${1.257}$ &    &$^{113}$Sn & ${1.168}$ &   &${0.9295}$&     \\
$^{114}$Sn & ${1.300}$ &    &$^{115}$Sn & ${1.312}$ &   &${1.0089}$&     \\
$^{116}$Sn & ${1.294}$ &    &$^{117}$Sn & ${1.279}$ &   &${0.9887}$&     \\
$^{118}$Sn & ${1.230}$ &1.237&$^{119}$Sn & ${1.220}$ &1.179&${0.9921}$&0.953\\
$^{120}$Sn & ${1.171}$ &1.190&$^{121}$Sn & ${1.151}$ &1.083&${0.9827}$&0.910\\
$^{122}$Sn & ${1.141}$ &1.103&$^{123}$Sn & ${1.107}$ &1.043&${0.9706}$&0.946\\
$^{124}$Sn & ${1.132}$ &1.047&$^{125}$Sn & ${1.088}$ &0.924&${0.9614}$&0.883\\
\end{tabular}
\end{ruledtabular}
\end{table*}

%%%%%%%%%%%%%%%%%%%%%%%%%%%%%%%%%%%%%%%%%%%%%%%%%%%%%%%%%%%%%%%%%%%%%%%%%%%%%%%%%%%%%
\begin{figure}[!ht]
\includegraphics[width=9cm,height=8cm]{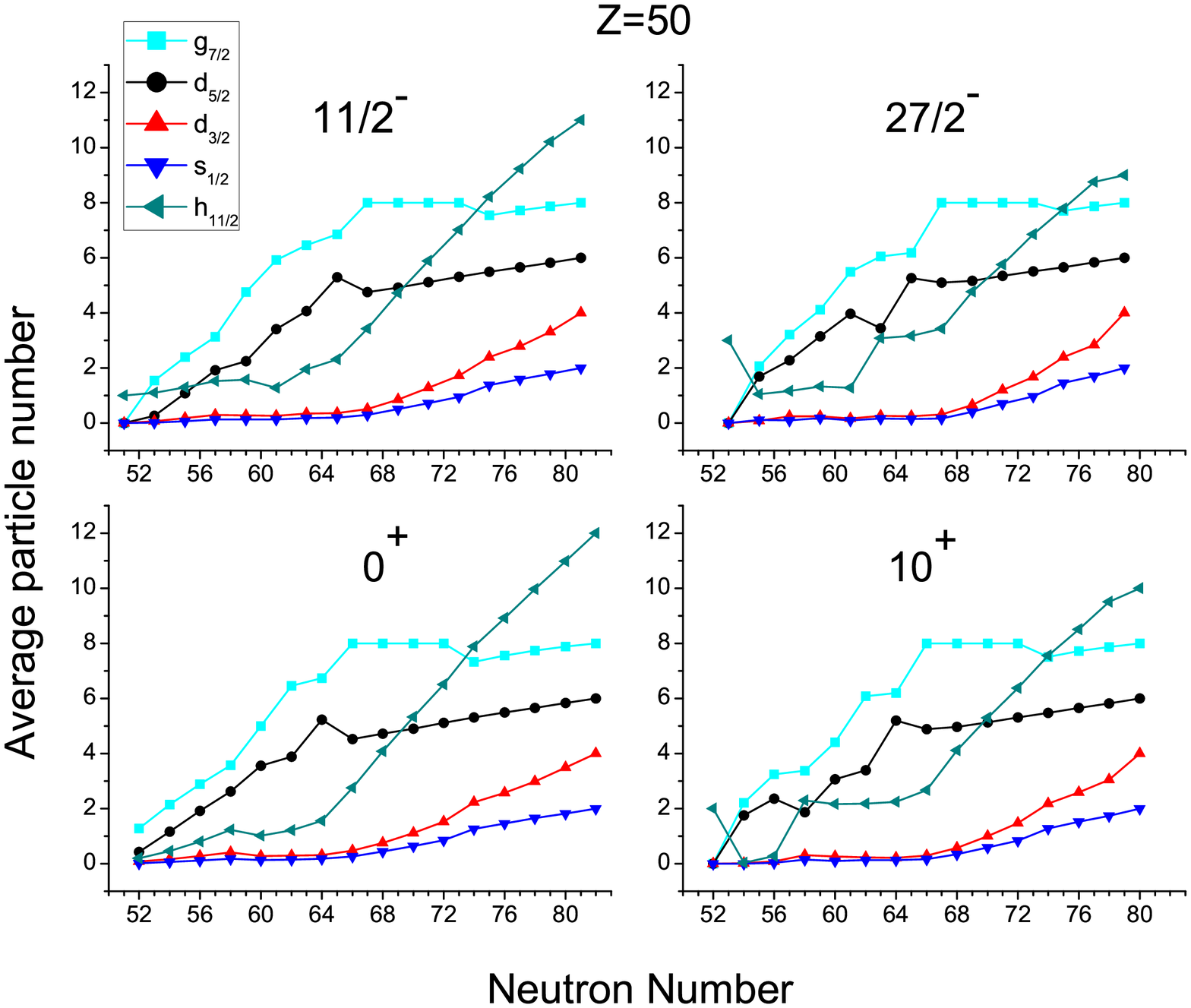}% Here is how to import EPS art
\caption{\label{fig:fig7}(Color online) The average particle number variation for different orbitals
with neutron number for the ${11/2}^-$, ${27/2}^-$ and ${0}^+$, ${10}^+$ states in the odd-A and even-even $Z$=50 isotopes respectively.}
\end{figure}
%%%%%%%%%%%%%%%%%%%%%%%%%%%%%%%%%%%%%%%%%%%%%%%%%%%%%%%%%%%%%%%%%%%%%%%%%%%%%%%%%%%%%
\begin{figure}[!ht]
\includegraphics[width=9cm,height=8cm]{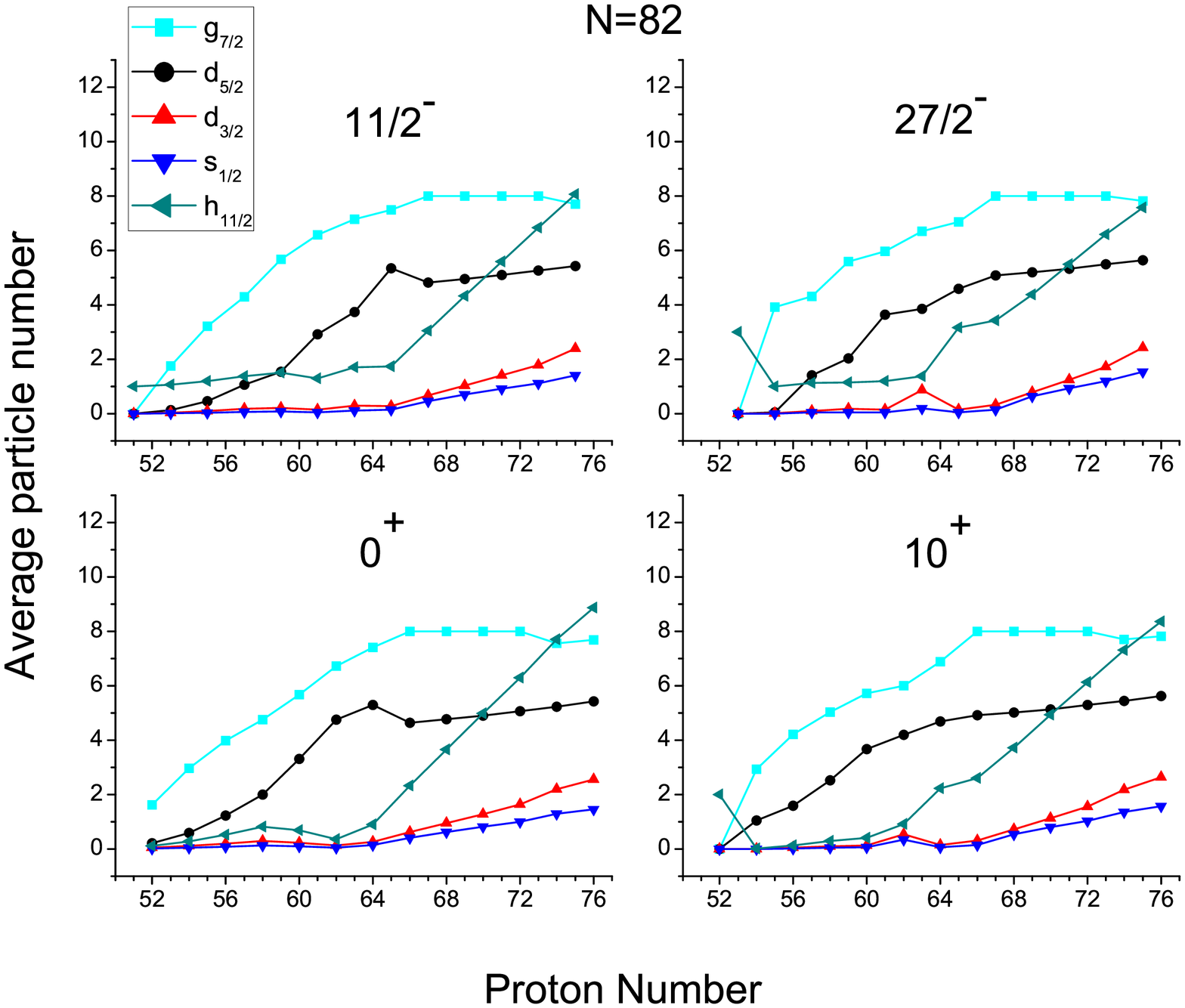}% Here is how to import EPS art
\caption{\label{fig:fig8}(Color online) The average particle number variation for different orbitals
with proton number for the ${11/2}^-$, ${27/2}^-$ and ${0}^+$, ${10}^+$ states in the odd-A and even-even $N$=82 isotones respectively.}
\end{figure}
%%%%%%%%%%%%%%%%%%%%%%%%%%%%%%%%%%%%%%%%%%%%%%%%%%%%%%%%%%%%%%%%%%%%%%%%%%%%%%%%%%%%%

\section{\label{sec:level3}Shell Model Calculations}
We have carried out large scale shell model calculations to understand the various features of both the isomeric chains. The calculations involving many identical nucleons in a single-j configuration are easier to deal with in the nuclear shell model~\cite{talmi93, lawson80, talmi03}, which is the case of isomers after the mid-shell. However, the calculations before the mid-shell involve identical nucleons in many-j orbitals. This results in the configuration mixing of different orbitals depending on their relative positions and the interaction matrix elements. 

\subsection{\label{sec:level3.1}Details of the calculations}
Neutrons in the $Z=50$ isotopes and protons in the $N=82$ isotones occupy the same valence space consisting of 0$g_{7/2}$, 1$d_{5/2}$, 1$d_{3/2}$, 2$s_{1/2}$ and 0$h_{11/2}$ orbitals lying between the magic proton/neutron numbers 50 and 82. The shell model calculations have been performed by using the Nushell code~\cite{brown07} along with the SN100PN~\cite{brown05} interaction. The calculations for the Sn-isomers have been done from $N=51$ to $81$, while for the $N=82$ isomers, the calculations have been done only up to the $Z=76$, near the proton-drip line. The proton and neutron single particle energies have been taken as 0.8072, 1.5623, 3.3160, 3.2238, 3.6051 MeV and -10.6089, -10.2893, -8.7167, -8.6944, -8.8152 MeV for the available 0$g_{7/2}$, 1$d_{5/2}$, 1$d_{3/2}$, 2$s_{1/2}$ and 0$h_{11/2}$ valence orbitals, respectively. The scaling factor varies with mass as $A^{-1/3}$, a well-established empirical law~\cite{brown01}.

We have done the calculations with full valence space up to the valence particle / hole number 8, i.e. from the valence nucleon number 51 to 58, and from 74 to 81 for both the $Z=50$ and the $N=82$ chains. No core excitations were taken into account. We further introduce some constraints for the valence nucleon numbers 59 to 65, to keep the dimensions in the limit by fixing the number of particles in the $g_{7/2}$ and $d_{5/2}$ orbitals, so that the calculations remain tractable and could be done in a reasonable time. Although we already have 8 particles which could fill up the $g_{7/2}$ orbital completely, there are strong chances of mixing of the $g_{7/2}$ and $d_{5/2}$ orbitals. We have, therefore, allowed the $g_{7/2}$ orbital to be filled with $4-8$ particles and the $d_{5/2}$ orbital with $0-2$ particles. We have further truncated the valence space by filling the $g_{7/2}$ orbital completely after the mid-shell for the nucleon numbers 66 to 73. It reduces the dimensions of the calculations significantly without affecting the desired results. As an example, we present in Fig. 3, a comparison of the results obtained from the full space calculations, and the truncated ones by freezing the $g_{7/2}$ orbital, in the case of $^{124}$Sn. We find that the energy of the ${10}^+$ isomeric state gets reduced in the truncated results, without affecting the pattern of the levels. Therefore, the aim of the present paper can still be accomplished, even with the truncated results in the mid-shell region, where the dimensions are quite large. We, therefore, expect that the calculated energies in the full space would be slightly higher, bringing them closer to the experimental values.

We have also calculated the effective single particle energies (ESPE) for both the chains by using the relation~\cite{otsuka01},

\begin{equation}
E_{espe} = E_{j_{n}}+\sum_{j_n} {\bar{E}(j_n,j_n)}\hat{n}_{j_{n}}
\end{equation}

where $E_{j_{n}}$ and $\hat{n}_{j_{n}}$ denote the single-particle energies and number operator for neutrons occupying the orbital $j_n$. The term is the monopole corrected interaction energy given by

\begin{equation}
{\bar{E}(j_n,j_n)} = \cfrac{\sum_J (2J+1) <j_n j_n; J|V|j_n j_n; J>}{\sum_J (2J+1)}
\end{equation}
 
where $<j_n j_n; J|V|j_n j_n; J>$ stands for a two-body matrix element of the effective interaction. We have plotted the ESPE's in the top and the bottom panels of Fig.~\ref{fig:fig4} for all the valence orbitals in the $Z=50$ and $N=82$ chains, respectively. We find that the $h_{11/2}$ orbital in both the chains has different relative positions (Fig.~\ref{fig:fig4}). Due to this, the nucleus $^{146}$Gd behaves like a doubly magic nucleus in the $N=82$ chain, whereas $^{114}$Sn does not in the $Z=50$ chain, which confirms the previous claims~\cite{ogawa78, kleinheinz79}. After the mid shell, the role of the $g_{7/2}$ and $d_{5/2}$ is rather neutralized, and the $h_{11/2}$ orbital begins to dominate.

\subsection{\label{sec:level3.2}Calculated systematics}

We present the calculated excitation energies for the $Z=50$ and $N=82$ isomeric chains in the bottom panels of Fig.~\ref{fig:fig1} and ~\ref{fig:fig2}, respectively. We find that all the three ${11/2}^-$, ${10}^+$ and ${27/2}^-$ isomeric states are reproduced reasonably well and follow the same systematic trends as the experimental data and a transition in the excitation energy is noticed near the mid-shell, which is more sharp in the $N=82$ chain at $Z=64$, due to the presence of the well known shell closure at $^{146}$Gd. The calculated energy gap between the ${11/2}^-$ and ${27/2}^-$ states, and the $0^+$ and ${10}^+$ states reproduces the experimental gap of $\sim$ 4 MeV from the nucleon number 51 to 58 and of $\sim$ 3 MeV from the nucleon number 74 to 81, where the full space calculations for the odd-A and even-A systems in both the $Z=50$ and $N=82$ chains could be carried out. The observed systematics are, therefore, reproduced quite well.

On the other hand, the energy gap is systematically smaller than the experimental values for the nucleon numbers $59-77$, a consequence of the truncations imposed by us. It declines from $\sim$ 4 MeV to $\sim$ 2 MeV in going from 59 to 65, and then becomes constant at $\sim$ 2 MeV, up to the nucleon number 73. As shown in Fig.~\ref{fig:fig3}, the truncation results in a smaller gap. However, it does not affect the aim of the present calculations as the ${10}^+$ and ${27/2}^-$ states closely follow each other in a very smooth manner even in the truncated region. It also suggests that similar structural change is responsible for the ${10}^+$ and ${27/2}^-$ states, relative to the $0^+$ and ${11/2}^-$ states respectively. 

The ${10}^+$ and ${27/2}^-$ isomers have nearly particle number independent energy before the mid-shell, even with the mixing of the $g_{7/2}$ and $d_{5/2}$ orbitals; we explain this in terms of the validity of the generalized seniority [10]. It was shown by Shlomo and Talmi~\cite{shlomo72} that the energy gap, between the state with generalized seniority $\it{v}$ (even) and the ground state with generalized seniority $\it{v}$=0 in a $2n+\it{v}$ nucleon system, denoted as $E((2n+\it{v})$ $\rightarrow$ $2n)$ is given by~\cite{shlomo72, talmi93}

\begin{eqnarray}
E((2n+\it{v})\rightarrow 2 n)=(E(\it{v},J)+E_n+\frac{1}{2} n\it{v}W)\nonumber\\
	-E_{n+\frac{1}{2} \it{v}}=E(\it{v},J)-E_{\frac{1}{2} \it{v}}
\end{eqnarray}
 
which is independent of $n$, the number of pairs. This is also equal to the energy gap between the energies of the two states in the $\it{v}$ nucleon system when ${n=0}$, as shown in the last equality. $E_n$ and $E(\it{v}, J)$ are the energy eigen-values for the $2n$ and $\it{v}$ nucleons with the generalized seniority 0 and $\it{v}$, respectively. W is a constant. Therefore, the ${10}^+$ states in the even-even nuclei before the mid-shell, have good generalized seniority. In the present work, we find that the energy gap between the ${27/2}^-$ states and the ${11/2}^-$ states in the odd-A isomers, also remains nearly constant and particle number independent, and equals to the gap between the ${10}^+$ states and the $0^+$ states in the even-A nuclei, before the mid-shell. This again suggests that the ${10}^+$ and ${27/2}^-$ states have similar structure except for the one extra nucleon in the $h_{11/2}$ orbital. Therefore, one may be assured about the validity of the generalized seniority, also in the odd-A nuclei, particularly for the ${27/2}^-$ states as the same j-orbital is involved in generating the ${27/2}^-$ and ${11/2}^-$ states, relative to the ${10}^+$ and $0^+$ states. 

Further, the situation becomes simpler after the mid-shell due to the dominant role of the unique parity $h_{11/2}$ orbital, which leads to a constant and particle number independent energy for ${10}^+$ and ${27/2}^-$ isomers, a signature of the seniority conservation~\cite{talmi93, lawson80, talmi03}. Thus, the empirical evidence supports good seniority for j-value as high as j=11/2. The seniority, therefore, remains approximately good for these isomers throughout the chain, with a structural transition near the mid-shell from generalized seniority regime to seniority regime.

%%%%%%%%%%%%%%%%%%%%%%%%%%%%%%%%%%%%%%%%%%%%%%%%%%%%%%%%%%%%%%%%%%%%%%%%%%%%%%%%%%%%%%%%%%%%%%%
\begin{table*}[htb]
\caption{\label{tab:table4}Same as Table III but for the $N=82$ isotones with Z $\geq$ 64. All the energies are in MeV.}
\begin{ruledtabular}
\begin{tabular}{c c c c c c c c}
Isotope & $ \Delta E^{2^+}_{0^+} $ & $\Delta E^{12^+}_{10^+}$ & Isotope & $\Delta E^{{15/2}^-}_{{11/2}^-}$ & $\Delta E^{{31/2}^-}_{{27/2}^-}$ & $R(15:2)$ & $R(31:12)$ \\
\hline
&&&&\\
$^{146}$Gd & ${1.972}$ &    &$^{147}$Tb & ${1.937}$ &   &${0.982}$&     \\
$^{148}$Dy & ${1.677}$ &1.932&$^{149}$Ho & ${1.560}$ &   &${0.930}$&     \\
$^{150}$Er & ${1.578}$ &1.446&$^{151}$Tm & ${1.478}$ &1.332&${0.937}$&0.921\\
$^{152}$Yb & ${1.531}$ &     &$^{153}$Lu & ${1.432}$ &    &${0.935}$&\\
\end{tabular}
\end{ruledtabular}
\end{table*}
%%%%%%%%%%%%%%%%%%%%%%%%%%%%%%%%%%%%%%%%%%%%%%%%%%%%%%%%%%%%%%%%%%%%%%%%%%%%%%%%%%%%%%%%%%%%%%%%
\begin{table*}[htb]
\caption{\label{tab:table5}Same as Table III, but calculated values only. Comparison of the experimentally measured $\Delta E^{2^+}_{0^+}$ and $ \Delta E^{{15/2}^-}_{{11/2}^-}$ $\gamma-$transitions in even-even and odd-A Sn-isotopes for $N$ $\ge$ 64. Also, compared are the $\Delta E^{12^+}_{10^+}$ and $ \Delta E^{{31/2}^-}_{{27/2}^-}$ $\gamma-$transitions involving states which decay to the ${10}^+$, ${27/2}^-$ isomers. All the energies are in MeV.}
\begin{ruledtabular}
\begin{tabular}{c c c c c c c c}
Isotope & $ \Delta E^{2^+}_{0^+} $ & $\Delta E^{{12}^+}_{{10}^+}$ & Isotope & $\Delta E^{{15/2}^-}_{{11/2}^-}$ & $\Delta E^{{31/2}^-}_{{27/2}^-}$ & $R(15:2)$ & $R(31:12)$ \\
\hline
&&&&&&&\\
$^{114}$Sn & ${1.508}$ & ${0.853}$ & $^{115}$Sn & ${1.463}$ & ${0.782}$ &${0.970}$ & ${0.916}$ \\
$^{116}$Sn & ${1.067}$ & ${1.688}$ & $^{117}$Sn & ${1.031}$ & ${1.698}$ &${0.966}$ & ${1.006}$ \\
$^{118}$Sn & ${0.989}$ & ${1.006}$ & $^{119}$Sn & ${0.943}$ & ${1.010}$ &${0.953}$ & ${1.004}$ \\
$^{120}$Sn & ${0.939}$ & ${0.906}$ & $^{121}$Sn & ${0.872}$ & ${0.871}$ &${0.928}$ & ${0.961}$ \\
$^{122}$Sn & ${0.887}$ & ${0.821}$ & $^{123}$Sn & ${0.826}$ & ${0.839}$ &${0.931}$ & ${1.022}$ \\
$^{124}$Sn & ${1.093}$ & ${0.937}$ & $^{125}$Sn & ${0.994}$ & ${0.905}$ &${0.910}$ & ${0.966}$ \\
$^{126}$Sn & ${1.123}$ & ${0.919}$ & $^{127}$Sn & ${1.014}$ & ${0.871}$ &${0.903}$ & ${0.948}$ \\
$^{128}$Sn & ${1.197}$ & ${0.977}$ & $^{129}$Sn & ${1.152}$ & ${}$ &${0.962}$ & ${ }$ \\
\end{tabular}
\end{ruledtabular}
\end{table*}
%%%%%%%%%%%%%%%%%%%%%%%%%%%%%%%%%%%%%%%%%%%%%%%%%%%%%%%%%%%%%%%%%%%%%%%%%%%%%%%%%%%%%%%%%%%%%%%%%%
%%%%%%%%%%%%%%%%%%%%%%%%%%%%%%%%%%%%%%%%%%%%%%%%%%%%%%%%%%%%%%%%%%%%%%%%%%%%%%%%%%%%%
\begin{table*}[htb]
\caption{\label{tab:table6} Same as Table V, but in the $N$=82 isotones for $Z$ $\ge$ 64. All the energies are in MeV.}
\begin{ruledtabular}
\begin{tabular}{c c c c c c c c}
Isotone & $ \Delta E^{2^+}_{0^+} $ & $\Delta E^{{12}^+}_{{10}^+}$ & Isotone & $\Delta E^{{15/2}^-}_{{11/2}^-}$ & $\Delta E^{{31/2}^-}_{{27/2}^-}$ & $R(15:2)$ & $R(31:12)$ \\
\hline
&&&&&&&\\
$^{146}$Gd & ${2.212}$ & ${0.994}$ & $^{147}$Tb & ${2.152}$ & ${0.920}$ & ${0.972}$ & ${0.925}$ \\
$^{148}$Dy & ${1.273}$ & ${1.626}$ & $^{149}$Ho & ${1.182}$ & ${1.501}$ & ${1.086}$ & ${0.923}$ \\
$^{150}$Er & ${1.162}$ & ${1.091}$ & $^{151}$Tm & ${1.067}$ & ${1.000}$ & ${0.918}$ & ${0.917}$ \\
$^{152}$Yb & ${1.102}$ & ${0.981}$ & $^{153}$Lu & ${1.006}$ & ${0.902}$ & ${0.913}$ & ${0.919}$ \\
$^{154}$Hf & ${1.068}$ & ${0.923}$ & $^{155}$Ta & ${0.974}$ & ${0.845}$ & ${0.912}$ & ${0.915}$ \\
$^{156}$W  & ${1.279}$ & ${1.035}$ & $^{157}$Re & ${1.135}$ & ${0.847}$ & ${0.887}$ & ${0.818}$ \\
$^{158}$Os & ${1.279}$ & ${1.002}$ & &&&&\\
\end{tabular}
\end{ruledtabular}
\end{table*}
%%%%%%%%%%%%%%%%%%%%%%%%%%%%%%%%%%%%%%%%%%%%%%%%%%%%%%%%%%%%%%%%%%%%%%%%%%%%%%%%%%%%%%%%%%%%%

\section{\label{sec:level4}Results and discussion}
\subsection{\label{sec:level4.1}The wave functions and the configuration assignments}

The maximum partition of each wave-function obtained from the shell model calculations have been presented in the form of bar charts for the Sn-isotopes and the $N=82$ isotones in Fig. 5 and 6, respectively. The bottom panels of Fig.~\ref{fig:fig5} and ~\ref{fig:fig6}, represent the occupancies related to the maximum partition of the wave functions for the $0^+$ and ${11/2}^-$ states in the $Z=50$ isotopic and the $N=82$ isotonic chains, respectively. On the other hand, the top panels of both the figures show the same for the ${10}^+$ and ${27/2}^-$ states. One can easily read the same by following the color coding. For example, the red color corresponds to the $d_{3/2}$ orbital. Its absence in a bar implies no particle in the $d_{3/2}$ orbital. Thus the final configuration related to the maximum partition of the wave function for the ${10}^+$ state of $^{122}$Sn-can be read from Fig.~\ref{fig:fig5} as {$g_{7/2}^8$ $d_{5/2}^6$ $d_{3/2}^2$ $s_{1/2}^0$ $h_{11/2}^6$}. 

We note that the nature of the wave function corresponding to the maximum partition of the $0^+$ ground state of an even-even Sn-isotope/ $N=82$ isotone is very similar to the ${11/2}^-$ state in the neighboring odd-A Sn-isotope/ $N=82$ isotone, except for the presence of one extra neutron/ proton in the $h_{11/2}$ orbital. Same is true for the ${10}^+$ state in an even-even Sn-isotope/ $N=82$ isotone and the ${27/2}^-$ state in the neighboring odd-A Sn-isotope/ $N=82$ isotone. It is also noteworthy that these wave functions change very systematically by maintaining an order across the full chain of Sn-isotopes/ $N=82$ isotones. These similarities in the nature of the wave functions may be taken as the origin of the similar systematic features in both the chains of isomers.

\subsection{\label{sec:level4.2}The average particle numbers and the seniority assignments}

Further insight into the configuration and structure of the various isomeric states can be obtained by plotting the average particle numbers in each of the valence orbitals for the Sn-isotopes, see Fig.~\ref{fig:fig7}. The four sections of the graph show the variation of the average particle number with increasing neutron number for the $0^+$, ${10}^+$ states, and for the ${11/2}^-$, ${27/2}^-$ states, respectively. The average particle number for the $d_{3/2}$ and $s_{1/2}$ orbitals is almost zero up to $N=66$, i.e. the mid-shell; it rises, thereafter, and changes in a very smooth and regular way, making a negligible contribution to the desired spin states. There is a competition between the remaining three $g_{7/2}$, $d_{5/2}$ and $h_{11/2}$ orbitals, which play an important role in deciding the final configuration of the desired spin state. The average particle number of the $h_{11/2}$ orbital also increases linearly after the mid-shell, which indicates that the $h_{11/2}$ orbital dominates over other orbitals. Similar discussion holds for the $N=82$ isotonic chain also, see Fig.~\ref{fig:fig8}.

The average particle number variation of the orbitals involved in generating the $0^+$ and ${11/2}^-$ state corresponding to an even-even and the odd-A Sn-isotope/ $N=82$ isotone respectively, exhibits almost similar behavior. It also suggests that the odd neutron, in the odd-A Sn-isotopes/ $N=82$ isotones, is totally aligned. We may, therefore, infer the seniority $\it{v}$=1 for the yrast ${11/2}^-$ isomeric state, coming from the last odd-particle in the $h_{11/2}$, and the seniority $\it{v}$=0 for the $0^+$ state, i.e. pair-correlated state, throughout the $Z=50$ and $N=82$ chains. 

Similarly, the variation of the average particle number, for the ${10}^+$ and ${27/2}^-$ isomeric states, is also observed to be similar in nature for both the chains. From the plots and the structure of the wave functions, we conclude that the ${27/2}^-$ isomeric state up to $N=61$ and $Z=63$ (except at N or $Z=53$) originates from one neutron/ proton in the $h_{11/2}$ orbital, and two broken pairs in the $(g_{7/2}, d_{5/2})$ orbitals, i.e. the generalized seniority becomes $\it{v}$=5, in both the $Z=50$ isotopes and $N=82$ isotones, respectively. We may extend the same argument for the ${10}^+$ isomeric states. It is seen to have the $h_{11/2}^0$ configuration, the total spin coming from the breaking of two neutron pairs in the $(g_{7/2}, d_{5/2})$ orbitals, i.e. the generalized seniority becomes $\it{v}$=4 for the $N=54-56$ isotopes in the $Z=50$ chain, and up to $Z=62$ in the $N=82$ chain. This happens due to the different scheme of the effective single particle energies for the protons as compared to the neutrons (see Fig.~\ref{fig:fig4}).

The particles keep on adding to the $h_{11/2}$ orbital, as the $g_{7/2}$ and $d_{5/2}$ orbitals are almost full for $N>64$. The seniority of the ${10}^+$ and ${27/2}^-$ isomeric states, therefore, remains constant at $\it{v}$=2 and 3 for both the chains, after the mid-shell. Our studies fully support the recent seniority assignments by Astier et al.~\cite{astier132, astier125} for the ${10}^+$ and ${27/2}^-$ isomers in the Sn-isotopes from $N=69$ onwards. We also confirm the previous seniority assignments made by Astier et al.~\cite{astier126} for the $Z=54-58$, $N=82$ isotones, and the seniority assignments discussed by McNeil et al.~\cite{mcneill89} for the range $Z=66-72$, $N=82$ isotones. We may conclude that the change in seniority is constant at $\Delta\it{v}$=4 in going from ${11/2}^-$ to ${27/2}^-$ and from $0^+$ to ${10}^+$, before the mid-shell, which declines to $\Delta\it{v}$=2, after the mid-shell. The generalized seniority assignments for both the $Z=50$ and $N=82$ chains along with the unpaired nucleon configurations have been summarized in the Tables I and II respectively, up to the nucleon number 64; the seniority assignments remains constant after 64 and are not listed in the tables.
%%%%%%%%%%%%%%%%%%%%%%%%%%%%%%%%%%%%%%%%%%%%%%%%%%%%%%%%%%%%%%%%%%%%%%%%%%%%%%%%%%%%%%%%%%%%%%
\subsection{\label{sec:level4.3}Aligned nature of the $h_{11/2}$ protons/neutrons}

Further evidence for the seniority assignment of various isomers and the states involved in the decay to the isomers by $\gamma$-transitions, is obtained from the alignment considerations. For this purpose, we compare the $\gamma$-transition energies and their ratios in the even-even core and the neighboring odd-A isotopic/ isotonic chain of isomers, respectively.

\subsubsection{\label{sec:level4.3.1}Experimental evidence}

We have listed the experimental E2 gamma energies associated with the transitions $\Delta E^{2^+}_{0^+}$  and $\Delta E^{{15/2}^-}_{{11/2}^-}$ for the even-even and odd-A Sn-isotopes in Table III~\cite{jain15, astier132, astier130, nndc}. The ratio of these transitions denoted as $R(15:2) = \Delta E^{{15/2}^-}_{{11/2}^-}$ / $\Delta E^{2^+}_{0^+}$ is observed to be $\sim$ 1 for the $^{114-125}$Sn-isotopes. This suggests a complete alignment of the odd-neutron in the $h_{11/2}$ orbital, producing the ${11/2}^-$ spin state. This supports our earlier observation that the ${11/2}^-$ state in odd-A Sn-isotopes and the $0^+$ state in the neighboring even-even Sn-isotopes have great similarity in their wave functions.

Similarly, the observed gamma transitions $\Delta E^{{12}^+}_{{10}^+}$ and $\Delta E^{{31/2}^-}_{{27/2}^-}$ in even-even and odd-A Sn-isotopes have also been listed in Table III for $^{118-125}$Sn-isotopes~\cite{ astier132, astier130}. Fotiades et al.~\cite{fotiades11} have compared the almost identical energies and similar structure involved in the $\Delta E^{2^+}_{0^+}$  and $\Delta E^{{12}^+}_{{10}^+}$ $\gamma$-transitions within the same isotope for $^{116-126}$Sn, and suggested that the ${10}^+$ isomeric state comes from the two aligned neutrons in the $h_{11/2}$ orbital. We have calculated the ratio of the transitions in odd-A Sn-isotope and its even-even core Sn-isotope, denoted as $ R(31:12)= \Delta E^{{31/2}^-}_{{27/2}^-}$ / $\Delta E^{{12}^+}_{{10}^+}$, which is also observed to be $\sim$ 1. The known gamma transition energies for the $N=82$ isotones, have also been listed in Table IV [46]. The ratios $R(15:2)$ and $R(31:12)$ again have the value $\sim$ 1, wherever these could be obtained. 

It is obvious from the observed values that the seniority $\it{v}$=1 for the ${11/2}^-$ states.  We may also conclude that the ${10}^+$ and ${27/2}^-$ isomeric states, in both the chains, are maximally aligned decoupled states involving two and three-neutrons in the $h_{11/2}$ orbital, respectively. That is why the ${10}^+$ and ${27/2}^-$ states closely follow each other in energy without exhibiting any odd-even effect. We, therefore, confirm the seniority after the mid-shell as $\it{v}$=0 for the $0^+$ states, $\it{v}$=1 for the ${11/2}^-$ states, $\it{v}$=2 for the $2^+$ states and $\it{v}$=3 for the ${15/2}^-$ states. Similarly, we assign the seniority $\it{v}$=2 for the ${10}^+$ states, $\it{v}$=3 for the ${27/2}^-$ states, $\it{v}$=4 for the ${12}^+$ states, and $\it{v}$=5 for the ${31/2}^-$ states, after the mid-shell. The same seniority difference $\Delta\it{v}$=2 between the ${15/2}^-$ and ${11/2}^-$ states, and for the $2^+$ and $0^+$ states gives their corresponding ratio $R(15:2)$ as $\sim$ 1. The difference $\Delta\it{v}$=2 also holds for the ${31/2}^-$ and ${27/2}^-$ states, and for the ${12}^+$ and ${10}^+$ states, which makes the ratio $R(31:12)$ $\sim$ 1.

%%%%%%%%%%%%%%%%%%%%%%%%%%%%%%%%%%%%%%%%%%%%%%%%%%%%%%%%%%%%%%%%%%%%%%%%%%%%%%%%%%%%%
\begin{table*}[htb]
\caption{\label{tab:table7}Predictions of the ${11/2}^-$, ${10}^+$ and ${27/2}^-$ states which are likely to be isomers, for both the $Z=50$ and $N=82$ chains. All the energies are in MeV.}
\begin{ruledtabular}
\begin{tabular}{c c c c c c c c c}
${ }$& \multicolumn{2}{c}{${10}^+$} & {} & \multicolumn{2}{c}{${11/2}^-$} & {} & \multicolumn{2}{c}{${27/2}^-$} \\
\hline\hline
Isotope &  Energy  &  Half-life & Isotope & Energy & Half-life & Isotope & Energy & Half-life \\
\hline
&&&&&&&&\\
$^{108}$Sn & ${\sim 4-4.5}$ & ${\sim 10-200}ps$ & $^{137}$Cs & ${\sim 2-2.5}$ & ${\sim 10-100}ps$ & $^{115}$Sn &${\sim 4.5-5}$ & ${\sim 0.1-10}ns$ \\

$^{110}$Sn & ${\sim 4-4.5}$ & ${\sim 10-200}ps$ & $^{139}$La & ${\sim 1.5-2}$ & ${\sim 10-100}ps$ & $^{117}$Sn &${\sim 3.5-4}$ & ${\sim 1-10}ns$ \\

$^{112}$Sn & ${\sim 4-4.5}$ & ${\sim 10-200}ps$ &  &  &  &$^{147}$Tb &${\sim 4}$ & ${\sim 1-20}ns$ \\

$^{136}$Xe & ${\sim 3.5-4}$ & ${\sim 100-500}ps$ &  &  &  & & &  \\

$^{144}$Sm & ${\sim 4-4.5}$ & ${\sim 100-500}ps$ &  &  &  & & &  \\

\end{tabular}
\end{ruledtabular}
\end{table*}
%%%%%%%%%%%%%%%%%%%%%%%%%%%%%%%%%%%%%%%%%%%%%%%%%%%%%%%%%%%%%%%%%%%%%%%%%%%%%%%%%%%%%%%%%%%%%%
%%%%%%%%%%%%%%%%%%%%%%%%%%%%%%%%%%%%%%%%%%%%%%%%%%%%%%%%%%%%%%%%%%%%%%%%%%%%%%%%%%%%%%%%%%%%%%
\begin{table*}[htb]
\caption{\label{tab:table8}A list of the known ${11/2}^-$, ${10}^+$ and ${27/2}^-$ states from the available experimental data~\cite{nndc} which are likely candidates for the isomeric states on the basis of our systematic studies. All the energies are in MeV.}
\begin{ruledtabular}
\begin{tabular}{c c c c c c c c c}
 &\multicolumn{3}{c}{$Z$=50} &  &\multicolumn{3}{c}{$N$=82}& \\
\hline\hline
 &Isotope &  Spin-parity  &  Energy& & Isotone & Spin-parity & Energy& \\
\hline
 &&&&&&&&\\
 &$^{108}$Sn& ${10}^+$ & {4.256} & &$^{136}$Xe & ${10}^+$ & 3.485&\\
 &$^{110}$Sn& $(10)$ & {4.317} & &$^{139}$La & ${11/2}^-$ & 1.420&\\
 &$^{112}$Sn& ${10}^+$ & 4.680 & &$^{147}$Tb & $({27/2}^-)$ & 3.889&\\
 &$^{115}$Sn& ${27/2}^-$ & 4.866 & &&&&\\
 &$^{117}$Sn& $({27/2}^-)$ & 3.824 & &&&&\\
\end{tabular}
\end{ruledtabular}
\end{table*}
%%%%%%%%%%%%%%%%%%%%%%%%%%%%%%%%%%%%%%%%%%%%%%%%%%%%%%%%%%%%%%%%%%%%%%%%%%%%%%%%%%%%%%%%%%%%%%
\subsubsection{\label{sec:level4.3.2}Comparison with the calculated results}
We have tabulated in Table V, the E2 gammas associated with the transitions $\Delta E^{2^+}_{0^+}$, $\Delta E^{{15/2}^-}_{{11/2}^-}$ and $\Delta E^{{12}^+}_{{10}^+}$, $\Delta E^{{31/2}^-}_{{27/2}^-}$ in the even-even and odd-A Sn-isotopes respectively, as obtained from the shell model calculations. We note that the calculated ratios $R(15:2)$ and $R(31:12)$ also remain $\sim$ 1. The calculations also suggest that the one, two, and three $h_{11/2}$ neutrons align to produce the ${11/2}^-$, ${10}^+$ and ${27/2}^-$ isomeric states respectively, after the mid-shell. Due to this, the ${10}^+$ and ${27/2}^-$ states closely follow each other after the mid-shell. 

We also summarize the calculated gammas $\Delta E^{2^+}_{0^+}$, $\Delta E^{{15/2}^-}_{{11/2}^-}$ and $\Delta E^{{12}^+}_{{10}^+}$, $\Delta E^{{31/2}^-}_{{27/2}^-}$ in the even-even and odd-A $N=82$ isotonic chain in Table VI. The ratios $R(15:2)$ and $R(31:27)$ again become $\sim$ 1. The alignment of $h_{11/2}$ protons takes place after the mid-shell, similar to the $h_{11/2}$ neutrons in the Sn-isotopic chain. We may, therefore, draw the same conclusions for the $N=82$ isotonic chain as for the $Z=50$ isotopic chain. Similar configurations and seniorities lead to similar alignment of the $h_{11/2}$ neutrons/protons for the ${11/2}^-$, ${10}^+$ and ${27/2}^-$ isomeric states in the both the $Z=50$ and $N=82$ chains. From these discussions, we may conclude that the isomeric states after the mid-shell are maximally aligned decoupled states.

\subsection{\label{sec:level4.4}Complete systematics of the even-even Sn-isotopes}

We present a complete overview of the experimental excitation energies for the yrast $0^+$, $2^+$, $4^+$, $6^+$, $8^+$ and ${10}^+$ states in the even-even Sn-isotopes from the neutron number 52 to 82, in Fig.~\ref{fig:fig10}. We can see that the energy of the $0^+$, $2^+$, $4^+$ and $6^+$ states is particle number independent throughout the chain due to same seniority $\it{v}$=2, except near the mid-shell, i.e. $N=64-66$, where the excitation energy shows a kink. This is due to a change in the orbitals involved in generating the states. The $g_{7/2}$ and $d_{5/2}$ orbitals hand over their role to the $h_{11/2}$ orbital near the mid-shell. The maximum spin-parity, which can be generated through the $g_{7/2}$ and $d_{5/2}$ orbitals, is $6^+$ from the seniority $\it{v}$=2 configuration. Therefore, the $8^+$ and ${10}^+$ spin-states require one more pair break-up before the $N=64$ nucleon number. The $8^+$ and ${10}^+$ states thus lie higher in energy before the mid-shell. They exhibit a transition to lower energy near $N=64-66$, which then becomes almost constant from $N=68$ to 80, since the $h_{11/2}$ orbital, with two unpaired nucleons, can generate the maximum ${10}^+$ spin-state without additional pair break up. The systematics, presented in Fig.~\ref{fig:fig10}, indicate that all the states from $0^+$ to ${10}^+$ undergo a transition from generalized seniority regime to seniority regime near the mid-shell.
%%%%%%%%%%%%%%%%%%%%%%%%%%%%%%%%%%%%%%%%%%%%%%%%%%%%%%%%%%%%%%%%%%%%%%%%%%%%%%%%%%%%%%%%%%%%%%
\begin{figure}[!ht]
\includegraphics[width=8cm,height=7cm]{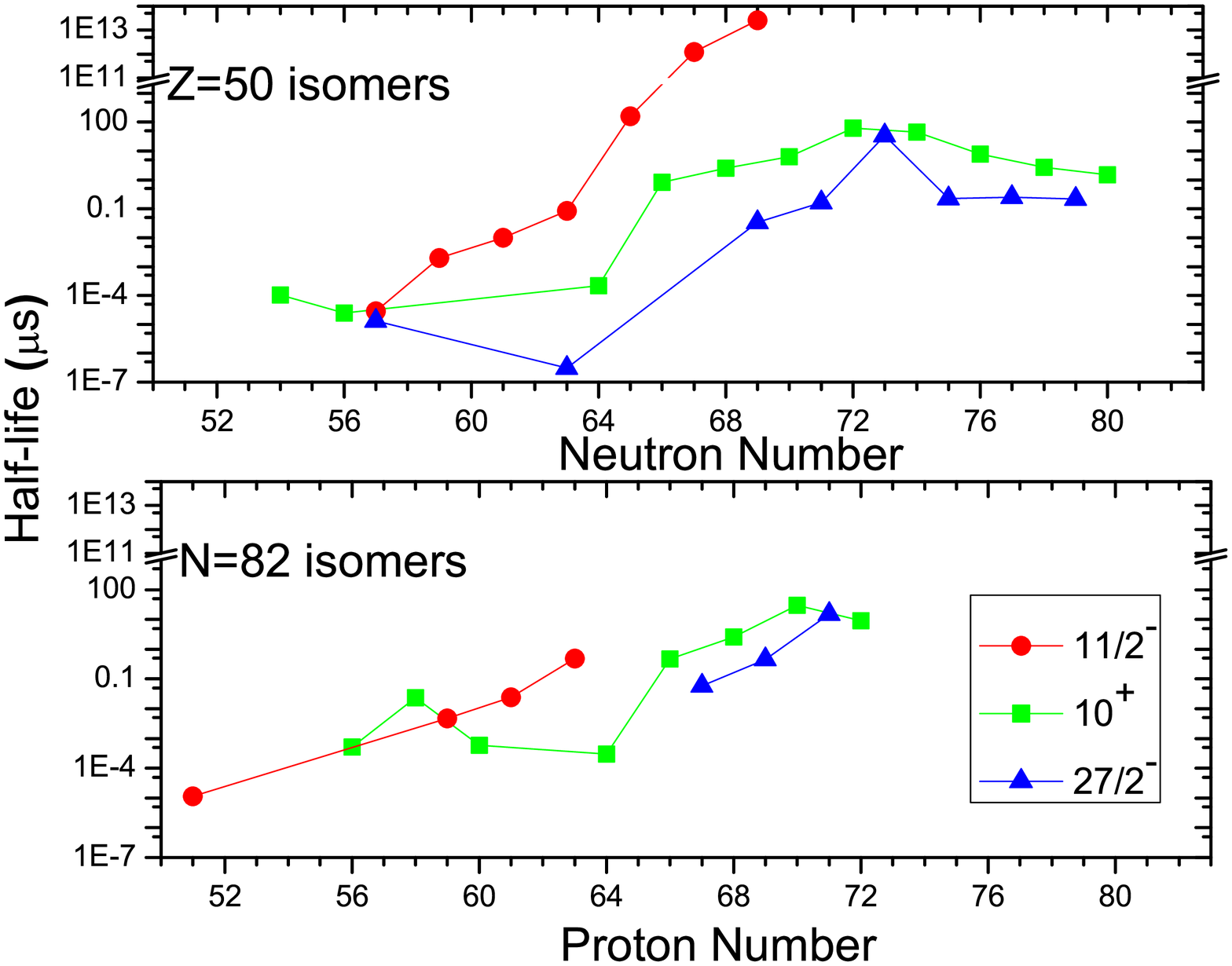}% Here is how to import EPS art
\caption{\label{fig:fig9}(Color online) Comparison of the experimental half-lives for $Z=50$ and $N=82$ isomers.}
\end{figure}
%%%%%%%%%%%%%%%%%%%%%%%%%%%%%%%%%%%%%%%%%%%%%%%%%%%%%%%%%%%%%%%%%%%%%%%%%%%%%%%%%%%%%%%%%%%%%%
\begin{figure}[!ht]
\includegraphics[width=9cm,height=7cm]{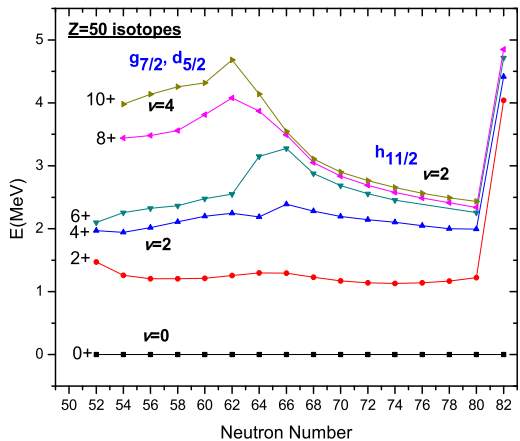}% Here is how to import EPS art
\caption{\label{fig:fig10}(Color online) Experimental energy systematics of the $0^+$ to ${10}^+$ yrast states in the even-even Sn-isotopes. The states are labeled by the seniority and the active orbitals.}
\end{figure}
%%%%%%%%%%%%%%%%%%%%%%%%%%%%%%%%%%%%%%%%%%%%%%%%%%%%%%%%%%%%%%%%%%%%%%%%%%%%%%%%%%%%%

\section{\label{sec:level5}Role of seniority on the half-life systematics}

We have plotted the measured half-lives (in $\mu$s) of these isomers with increasing nucleon numbers in Fig.~\ref{fig:fig9}. Due to the wide range of half-lives, we have plotted the log values of the half-lives and, also introduced a break in the vertical scale in Fig.~\ref{fig:fig9}. The half-life systematics for the $Z=50$ isomers are plotted in the top panel, while those for the $N=82$ isomers are plotted in the bottom panel. The half-lives of the ${11/2}^-$ isomers show a similar trend with increasing nucleon number in both the chains, although valence protons are involved in one case, while valence neutrons are involved in the other. This may be attributed to similar structure of these states. As we increase the nucleon numbers, the gamma decay energy keeps falling and the multipolarity rises to M4. As a result, the half-lives increase. The decay mode changes at the peaks in the half-lives, and beta decay/EC comes into play; we, therefore, do not discuss these cases in the present work.

The half-lives of the ${10}^+$ and ${27/2}^-$ isomers exhibit a rise near the half-filled $h_{11/2}$ configuration, attain a maximum value, and fall with increasing nucleon number. The ${10}^+$ and ${27/2}^-$ isomeric states, for the $Z=50$ isomers, exhibit a maximum at the neutron numbers 72 and 73 respectively, where the $h_{11/2}$ neutron orbital becomes half-filled~\cite{mayer94, zhang00, lozeva08}. On the other hand, for the $N=82$ isomers, the peaks are observed at $Z=70$ and 71 for the ${10}^+$ and ${27/2}^-$ isomeric states respectively, where the $h_{11/2}$ proton orbital becomes half-filled~\cite{mcneill89}. This happens because the electric quadrupole (E2) transition between a state $J_i$ and another state $J_f$, both having same seniority $\it{v}$, in the $j^n$ configuration is given by~\cite{lawson80, talmi03, isacker10}

\begin{eqnarray}
B(E2)=\frac{1}{2J_i+1}|\frac{2j+1-2n}{2j+1-2\it{v}} <j^v v J_f|| \nonumber \\
\sum_i r_i^2 Y_2(\theta_i \phi_i)||j^v v J_i>|^2
\end{eqnarray}
 
When $n$ approaches $(2j+1)/2$, i.e. the half-filled configuration, the B(E2) vanishes. We, therefore, conclude that the isomeric half-lives at the half-filled $h_{11/2}$ orbital in the $Z=50$ and $N=82$ chains are most affected by the seniority selection rules. As we move away from the region of half-filled orbital, the role of seniority in slowing down the transitions diminishes. We also find that the half-life of the ${10}^+$ isomers in the even-even nuclei are $\sim$ 100 times larger than the half-life of the known ${27/2}^-$ isomers in the odd-A nuclei before the mid-shell. This may be a direct consequence of a larger seniority mixing in the ${27/2}^-$ states of the odd-A nuclei as compared to the ${10}^+$ states in even-even nuclei~\cite{talmi93, lawson80}. The situation becomes rather complex due to the multi-j shell configurations for these isomers in both the $Z=50$ isotopic and $N=82$ isotonic chains. 

\section{\label{sec:level6}Predictions of possible new isomers}

Similar configurations, occupancies and wave functions lead to similar trends in the excitation energy for both the $Z=50$ and $N=82$ chains. Therefore, a knowledge of basic key features of one region enables us to make predictions for the unknown isomers/isomeric properties in the other region. It is well known that the population of nuclei at the drip lines is rather difficult. The systematic features may, therefore, become a great tool to predict the isomers in the near the drip lines. We can see that the ${11/2}^-$, ${10}^+$ and ${27/2}^-$ isomeric states are observed with many gaps for the $Z=50$ isotopic/ $N=82$ isotonic chains within the same valence space $50-82$. The above results and discussion strongly suggest the existence of new isomers in the gaps, with almost similar excitation energies. We have summarized our predictions for the new isomers in Tables VII. We have not listed those cases where the predicted half-lives may be lower than 10 ps. These predictions may open the way for possible experiments to explore the new isomers. We have also listed the available experimental data~\cite{nndc} of the yrast ${11/2}^-$, ${10}^+$ and ${27/2}^-$ states with their excitation energies in the Table VIII, which are yet to be characterized as isomeric states. They closely follow the systematics presented in this paper. All of these states already appear in the predictions in Tables VII, and are good candidates for isomers which needs to be confirmed by life-time measurements.

\section{\label{sec:level7}Summary and Conclusions}

To summarize, we have presented a comprehensive overview of the excitation energy systematics of the ${11/2}^-$, ${10}^+$ and ${27/2}^-$ isomeric states, throughout the $Z=50$ isotopic and $N=82$ isotonic chains, which are observed to display nearly identical behavior. A constant and particle number independent energy gap of $\sim$ 4 MeV between the $0^+$ and ${10}^+$ states and, the ${11/2}^-$ and ${27/2}^-$ states, exists before the mid-shell, which becomes $\sim$ 3 MeV after the mid-shell. Large scale shell model calculations are able to reproduce the experimental trends reasonably well, except for the energy gap which is consistently smaller by more than 0.5 MeV in the region of truncations. The relative position of the $h_{11/2}$ orbital effective single particle energies in both the chains has been shown to be related to the magicity of $Z=64$ in the $N=82$ isotonic chain. 

We have obtained the configurations, wave functions and average particle numbers to understand their origin in terms of seniority. We find that configuration mixing exists before the mid-shell, due to the presence of many-j orbitals, but the isomers still show the empirical features of the seniority scheme. These have been understood in terms of generalized seniority. The ${10}^+$ and ${27/2}^-$ isomeric states appear to have good generalized seniority before the mid-shell, whereas seniority remains conserved after the mid-shell due to the dominant role of the $h_{11/2}$ orbital. The transition from the generalized seniority regime to the seniority regime is responsible for the change in the energy gap near the mid-shell in both the chains. 

We also find that the overall structure of the ${11/2}^-$ and ${27/2}^-$ states in the odd-A nuclei is very similar to the $0^+$ and ${10}^+$ states of the neighboring even-even nuclei, in both the chains. Generalized seniority in even-even nuclei has earlier been shown to be valid for low lying states; we find that it remains valid up to ${10}^+$ states. We have been able to extend the validity of genarlized seniority to odd-A nuclei before the mid-shell. 

By using these arguments and also the alignment properties of the $h_{11/2}$ nucleons after the mid-shell, we have been able to assign the seniority for the $2^+$, ${12}^+$, ${15/2}^-$ and ${31/2}^-$ states, which decay to the $0^+$, ${10}^+$, ${11/2}^-$ and ${27/2}^-$ states via E2 decay, respectively. This confirms a constant change in seniority of $\Delta\it{v}$=2 between the $0^+$ and ${10}^+$ states, and also between the ${11/2}^-$ and ${27/2}^-$ states after the mid-shell. We also conclude that the high-spin isomers behave as maximally aligned decoupled states after the mid-shell. 

We have also explained the similar half-life trends of the isomers in both the chains in terms of seniority. The seniority selection rule governs the behavior of the half-life near $N=72$ and $Z=70$ in the $Z=50$ and $N=82$ chains, respectively. Finally, we have presented a complete picture of the $0^+$ to ${10}^+$ yrast states of the even-even Sn-isotopes in terms of generalized seniority before the mid-shell, and seniority after the mid-shell. This agrees with our interpretation of the semi-magic isomers.

To conclude, the present work highlights the important role of seniority, generalized seniority and the alignment properties of the semi-magic isomers, which enable us to explain the similar features of the two semi-magic chains. We note that the Sn-isotopes approaching the neutron-drip line behave in a way very similar to the $N=82$ isotones approaching the proton-drip line. The similarity also holds to a significant extent for the neutron-deficient Sn-isotopes and the proton-deficient $N=82$ isotones. The similar overall trends in both the chains attest to the charge independent nature of the effective nuclear interactions over a wide range of isospin. These studies have also enabled us in making predictions of possible new isomers for future experiments.

%%%%%%%%%%%%%%%%%%%%%%%%%%%%%%%%%%%%%%%%%%%%%%%%%%%%%%%%%%%%%%%%%%%%%%%%%%%%%%%%%%%%%%%%%%
\begin{acknowledgments}
The authors thank Dr. P. C. Srivastava for his assistance in the initial phases of this work. One of the authors acknowledges the financial support from MHRD (Govt. of India) in the form of a fellowship. Financial support from the DST (Govt. of India) and DAE (Govt. of India) is also acknowledged.
\end{acknowledgments}

%%%%%%%%%%%%%%%%%%%%%%%%%%%%%%%%%%%%%%%%%%%%%%%%%%%%%%%%%%%%%%%%%%%%%%%%%%%%%%%%%%%%%%%%%%%
\bibliography{apssamp}% Produces the bibliography via BibTeX.

\begin{thebibliography}{50}

\bibitem{walker99}  P. Walker and G. Dracoulis, Nature $\textbf{399}$, 35 (1999).
\bibitem{jain15}	A.K. Jain, Bhoomika Maheshwari, Swati Garg, Monika Patial and Balraj Singh, Nuclear Data Sheets (accepted).
\bibitem{jain14}	A. K. Jain, Invited talk at the 6th Asian Nuclear Physics Association Symposium, VECC, Kolkata, 19-21 February, 2014; Invited talk at the International Workshop on Nuclear Theory, Rila Mountains, Bulgaria, 22-28 June, 2014.
\bibitem{racah42}	G. Racah, Phys. Rev. $\textbf{61}$, 186 (1942); Phys. Rev. $\textbf{63}$, 367 (1943).
\bibitem{talmi93}	I. Talmi, Simple Models of Complex Nuclei. The Shell Model and Interacting Boson Model (Harwood, Academic, Chur, Switzerland, 1993).
\bibitem{lawson80}	R. D. Lawson, Theory of the Nuclear Shell Model, (Oxford University Press, New York, 1980).
\bibitem{talmi03}	I. Talmi, Fifty years of the Shell Model ̶ The Quest for the Effective Interaction, Adv. Nucl. Phys. $\textbf{27}$, 1 (2003).
\bibitem{isacker10}	P. Van Isacker, AIP Conf. Proc. $\textbf{1323}$, 141 (2010).
\bibitem{isacker11}	P. Van Isacker, Journal of Physics: Conf. Series $\textbf{322}$, 012003 (2011).
\bibitem{talmi71}	I. Talmi, Nucl. Phys. A $\textbf{172}$, 1 (1971).
\bibitem{sandulescu97}	N. Sandulescu, J. Blomqvist, T. Engeland, M. Hjorth-Jensen, A. Holt, R. J. Liotaa and E. Osnes, Phys. Rev. C $\textbf{55}$, 5 (1997).
\bibitem{scholten83}	O. Scholten and H. Kruse, Phys. Lett. B $\textbf{125}$, 113 (1983).
\bibitem{daly80}	P. J. Daly, P. Klenheinz, R. Broda, S. Lunardi, H. Backe, and J. Blomqvist, Z. Phys. A $\textbf{298}$, 173 (1980).
\bibitem{fogelberg81}	B. Fogelberg, K. Heyde, J. Sau, Nucl. Phys. A $\textbf{352}$, 157 (1981).
\bibitem{daly86}	P. J. Daly et al., Z. Phys. A $\textbf{323}$, 245 (1986).
\bibitem{lunardi87}	S. Lunardi et al., Z. Phys. A $\textbf{328}$, 487 (1987).
\bibitem{broda92}	R. Broda et al., Phys. Rev. Lett. $\textbf{68}$, 1671 (1992).
\bibitem{mayer94}	R. Mayer et al., Phys. Lett. B $\textbf{336}$, 308 (1994).
\bibitem{daly95}	P. J. Daly, R. Mayer et al., Phys. Scr. T $\textbf{56}$, 94 (1995).
\bibitem{pinston00}	J. A. Pinston et al., Phys. Rev. C $\textbf{61}$, 024312 (2000).
\bibitem{zhang00}C. T. Zhang et al., Phys. Rev. C $\textbf{62}$, 057305 (2000).
\bibitem{lozeva08} R. L. Lozeva et al., Phys. Rev. C $\textbf{77}$, 064313 (2008).
\bibitem{pietri11}S. Pietri et al., Phys. Rev. C $\textbf{83}$, 044328 (2011).
\bibitem{astier132}A. Astier, Journal of Physics: Conf. Series $\textbf{420}$, 012055 (2013).
\bibitem{astier125}A. Astier et al., Phys. Rev. C $\textbf{85}$, 054316 (2012).
\bibitem{iskra14}L. W. Iskra et al., Phys. Rev. C $\textbf{89}$, 044324 (2014).
\bibitem{bron79}J. Bron et al., Nucl. Phys. A $\textbf{318}$, 335 (1979).
\bibitem{poelgeest80}A. Van. Poelgeest et al., Nucl. Phys. A $\textbf{346}$, 70 (1980).
\bibitem{harada88}	H. Harada et al., Phys. Lett. B $\textbf{207}$, 17 (1988).
\bibitem{savelius98}	A. Savelius et al., Nucl. Phys. A $\textbf{637}$, 491 (1998).
\bibitem{gableske01}	J. Gableske et al., Nucl. Phys. A $\textbf{691}$, 551 (2001).
\bibitem{wolinska05}	M. Wolinska-Cichocka et al., Eur. Phys. J. A $\textbf{24}$, 259 (2005).
\bibitem{wang10}	S. Y. Wang et al., Phys. Rev. C $\textbf{81}$, 017301 (2010).
\bibitem{fotiades11}	N. Fotiades et al., Phys. Rev. C $\textbf{84}$, 054310 (2011).
\bibitem{simpson14}	G. S. Simpson et al., Phys. Rev. Lett. $\textbf{113}$, 132502 (2014).
\bibitem{maheshwari15}	B. Maheshwari, A. K. Jain and P. C. Srivastava, Phys. Rev. C $\textbf{91}$, 024321 (2015).
\bibitem{mcneill89}	J. H. Mcneill et al., Phys. Rev. Lett. $\textbf{63}$, 860 (1989).
\bibitem{astier126}	A. Astier et al., Phys. Rev. C $\textbf{85}$, 064316 (2012).
\bibitem{brown07}	B. A Brown and W. D. M. Rae, Nushell@MSU, MSU-NSCL report (2007).
\bibitem{brown05}	B. A Brown, N. J. Stone, J. R. Stone, I. S. Towner, and M. Hjorth-Jensen, Phys. Rev. C $\textbf{71}$, 044317 (2005).
\bibitem{brown01}	B. A. Brown, Prog. In Part. And Nucl. Phys. $\textbf{47}$, 517 (2001).
\bibitem{otsuka01}	T. Otsuka, R. Fujimoto, Y. Utsuno, B. A. Brown, M. Honma, and T. Mizusaki, Phys. Rev. Lett. $\textbf{87}$, 082502 (2001).
\bibitem{ogawa78}	M. Ogawa et al., Phys. Rev. Lett. $\textbf{41}$, 289 (1978). 
\bibitem{kleinheinz79}	P. Kleinheinz et al., Z. Phys. A $\textbf{290}$, 279 (1979).
\bibitem{shlomo72}  S. Shlomo, I. Talmi, Nucl. Phys. A $\textbf{198}$, 81 (1972).
\bibitem{astier130}	A. Astier, EPJ Web of Conf. $\textbf{62}$, 01008 (2013).
\bibitem{nndc}	www.nndc.bnl.gov/ENSDF 
\end{thebibliography}

\end{document}